\documentclass[12pt,preprint]{aastex}
\usepackage{epsf,emapjbib_only,multicol}
\usepackage{epsfig}
\newcommand{\uv}{\mbox{$u$-$v$}}

\newcommand{\ka}{{30\,GHz}}
\newcommand{\midka}{{31\,{\rm GHz}}}

\newcommand{\gsim}{\gtrsim}

\newcommand{\degree}{^\circ}
\newcommand{\rms}{{\it rms}}

\def\clA{Cl\,J1415.1+3612}
\def\clB{Cl\,J1429.0+4241}
\def\clC{Cl\,J1226.9+3332}
\def\betaModel{$\beta$-model}

\begin{document}
\title{Observations of High-Redshift X-Ray Selected Clusters with the Sunyaev-Zel'dovich Array}

\author{Stephen~Muchovej$^{1,2}$,
John~E.~Carlstrom$^{3,4,5,6}$, John~Cartwright$^{3,4}$,
Christopher~Greer$^{3,4}$, David~Hawkins$^{7}$,
Ryan~Hennessy$^{3,4}$, Marshall~Joy$^{8}$,
James~W.~Lamb$^{7}$, Erik~M.~Leitch$^{9}$, Michael~Loh$^{3,6}$,
Amber~D.~Miller$^{1,10,11}$, Tony~Mroczkowski$^{1,2}$, Clem~Pryke$^{3,4,5}$,
Ben~Reddall$^{3}$, Marcus~Runyan$^{3,5}$,
Matthew~Sharp$^{3,6}$ and David~Woody$^{7}$}

\footnotetext[1]{Columbia Astrophysics Laboratory, Columbia University, New York, NY 10027}
\footnotetext[2]{Department of Astronomy, Columbia University, New York, NY 10027}
\footnotetext[3]{Kavli Institute for Cosmological Physics, University of Chicago, Chicago, IL 60637}
\footnotetext[4]{Department of Astronomy and Astrophysics, University of Chicago, Chicago, IL 60637}
\footnotetext[5]{Enrico Fermi Institute, University of Chicago, Chicago, IL 60637}
\footnotetext[6]{Department of Physics, University of Chicago, Chicago, IL 60637}
\footnotetext[7]{California Institute of Technology, Owens Valley Radio Observatory, Big Pine, CA 93513} 
\footnotetext[8]{Department of Space Science,VP62, NASA Marshall Space Flight Center, Huntsville, AL 35812}
\footnotetext[9]{\parbox[t]{6in}{NASA Jet Propulsion Laboratory, Pasadena, CA 91109; \\ Department of Astronomy, California Institute of Technology, Pasadena CA 91125}}
\footnotetext[10]{Department of Physics, Columbia University, New York, NY 10027}
\footnotetext[11]{Alfred P. Sloan Fellow}

\begin{abstract}
We report measurements of the Sunyaev-Zel'dovich (SZ) effect in three high-redshift
($0.89 \le z \le 1.03 $),
X-ray selected galaxy clusters. The observations were obtained
at 30 GHz during the commissioning period of a new,
eight-element interferometer -- the Sunyaev-Zel'dovich Array (SZA) -- built
for dedicated SZ effect observations.  
The SZA observations are sensitive to angular scales larger than
those subtended by the virial radii of the clusters. 
Assuming isothermality and hydrostatic equilibrium for the
intracluster medium, and gas-mass fractions consistent with those
for clusters at moderate redshift, we calculate electron temperatures, gas
masses, and total cluster masses from the SZ data.
The SZ-derived masses, integrated approximately to
the virial radii, are
$1.9 ^{+0.5}_{-0.4} \times 10^{14}M_\odot$
for \clA, 
$3.4 ^{+0.6}_{-0.5} \times 10^{14}M_\odot$
for \clB\ and
$7.2 ^{+1.3}_{-0.9} \times 10^{14}M_\odot$
for \clC.
The SZ-derived quantities are in good agreement
with the cluster properties derived from X-ray measurements. 
\end{abstract}

\keywords{cosmology: observations --- galaxies: individual (NGC5529) --- clusters: individual 
(\clA, \clB, \clC) --- Sunyaev-Zel'dovich Effect --- cosmic microwave
background --- techniques: interferometric}

\section{Introduction}
\label{sec:intro}

Galaxy clusters are the most massive, gravitationally-bound structures in the Universe. 
Over a Hubble time, they form from the rare, high-density peaks in the primordial density field 
on scales of  $\sim10~{\rm Mpc}$. As their abundance and 
evolution are critically dependent on cosmology, there is 
considerable
interest in finding clusters at high redshift ($z \ge 1$). To date,
only a few massive clusters at $z >1 $ have been identified
and confirmed, primarily through the detection of extended  X-ray emission
from the hot intracluster medium (ICM)
\citep[e.g.,][]{mullis2005,bremer2006,maughan2006,stanford2006}. 
In addition to
 X-ray observations,
optical and infrared observations of the cluster member galaxies,
and weak lensing of background galaxies by the deep
cluster potential, are 
 complementary probes  of high-redshift clusters
\citep[e.g.,][]{refregier2003,gladders2006,stanford2005,clowe2006,wittman2006}. 

Recently, Sunyaev-Zel'dovich effect measurements of galaxy clusters have 
emerged as a powerful probe of cluster physics and cosmology \citep[for a review, see][]{carlstrom2002}.
Measurements of the SZ effect have been used to 
determine cluster properties such as the gas and total masses, electron temperatures and scaling
relations, as well as to constrain $H_0$ and the cosmological distance scale \citep[e.g.,][]{hughes1998,mason2001,grego2001,reese2002,mccarthy2003,benson2004,jones2005,afshordi2005,laroque2006,bonamente2006}.
Several telescopes specifically designed for observations of the SZ effect 
in high-redshift clusters
are currently operating or under development, including
the Sunyaev-Zel'dovich Array, the Arcminute Microkelvin Imager
\citep{kaneko2006}, the Atacama Cosmology Telescope \citep{fowler2004} and
the South Pole Telescope \citep{ruhl04}.

The SZ effect is a spectral distortion of the cosmic microwave background (CMB) radiation caused by inverse
Compton scattering of the CMB photons by electrons in the hot ICM
(\citealt{sunyaev70,sunyaev72}, see also \citealt{birkinshaw1999}). The magnitude of the effect is proportional
to the integrated pressure of the ICM, i.e., the density of electrons along the line of sight, weighted
by the electron temperature. The SZ flux of a cluster is therefore
a measure of its total thermal energy.  

The change in the observed brightness of the CMB 
caused by the SZ effect is given by
\begin{equation}
\label{y}
\frac{\Delta T_{\rm CMB}}{T_{\rm CMB}}=f(x)\int \sigma_{\rm T}n_e\frac{k_BT_e}{m_ec^2} dl \equiv f(x)y
\end{equation}
where $T_{\rm CMB}$ is the cosmic microwave background temperature (2.73 K), 
$\sigma_{\rm T}$ is the Thomson scattering cross section, $k_B$ is Boltzmann's constant, $c$ is the speed of light,  
and $m_e$, $n_e$, and $T_e$ are the electron mass, number density and temperature. 
 Equation~(\ref{y}) defines the Compton $y$-parameter.
The frequency dependence of the SZ effect is contained in the term
\begin{equation}
\label{f}
f(x)=\left(x\frac{e^x+1}{e^x-1}-4\right)\left(1+\delta_{\rm SZ}(x,T_e)\right) ,
\end{equation}
where $x\equiv h\nu/{k_BT_{\rm CMB}}$, $h$ is Planck's constant, and $\delta_{\rm SZ}$ is a
relativistic correction, for which we adopt the \citet{itoh1998}
calculation, valid to fifth order in $k_BT_e/m_ec^2$. The SZ effect appears as a temperature decrement
at frequencies below $\approx 218~{\rm GHz}$, and an increment at higher frequencies.

The redshift independence of the SZ
effect in both brightness and frequency (the ratio $\Delta T/T$ in equation~(\ref{y}) is independent of the
distance to the cluster) offers enormous potential for finding
high-redshift clusters. A cluster catalog resulting from an SZ survey 
of uniform sensitivity will be limited
by a cluster mass threshold that is only  weakly dependent on redshift for
$z\gsim 0.5$ (via the angular diameter distance). To
realize the full potential of SZ surveys for cosmology
will require a thorough understanding of the relationship
between SZ observables and cluster mass, achievable through detailed
cluster observations.

The SZA is a new, eight-element array of 3.5-meter precision
telescopes designed to conduct SZ surveys 
at 30 GHz and detailed cluster observations at
30 GHz and 90 GHz. We present observations
obtained during the commissioning of the SZA, of 
three high-redshift,
X-ray selected clusters: ClJ1226.9+3332 at $z=0.89$, 
ClJ1429.0+4241 at $z=0.92$
and ClJ1415.1+3612 at $z=1.03$.  We describe the
analysis of the SZ data and compare the derived cluster properties
with those determined from X-ray observations.  We also compare the
SZA observations of ClJ1226.9+3332 with previous BIMA observations of the
SZ effect \citep{joy2001} in the same cluster. No previous SZ measurements of the other two clusters have been reported. 

The paper is organized as follows: In \S2 we describe the
Sunyaev-Zel'dovich Array, followed by a description of the data acquisition, 
reduction and calibration.  The data analysis is presented in \S3. 
In \S4 we present the results of the analysis and compare with previous 
SZ and X-ray results. Conclusions are given in \S5.

\section{Observations and Data Reduction}
\subsection{The Sunyaev-Zel'dovich Array}
\label{sec:obs}

The Sunyaev-Zel'dovich Array (SZA) is a new interferometer equipped
with sensitive centimeter (26--36 GHz) and millimeter-wavelength (80--115 GHz)
receivers, designed specifically for detecting and imaging the SZ effect in
galaxy clusters.  In this paper, we report results only from the
centimeter-wavelength (hereafter \ka )
SZA observations.

An interferometer has sensitivity to angular scales up to 
the resolution of its shortest baseline.  To provide a good match to
galaxy clusters, which at $z\gtrsim0.1$ subtend several arcminutes on the sky,
the SZA was designed with small (3.5 meter) antennas,
allowing close-packed configurations with sensitivity to scales as
large as $5\arcmin$ at 30 GHz.  The specific choice of diameter provides
optimum brightness sensitivity to the SZ effect from distant clusters, 
while filtering out the contamination by the intrinsic anisotropy of the CMB 
at larger angular scales.
The instantaneous field of view of the SZA is given by the primary
beam of the 3.5-meter antennas, approximately $12\arcmin$ (FWHM) at 30
GHz.

\begin{figure}[ht]
\begin{center}

\includegraphics[scale=0.8]{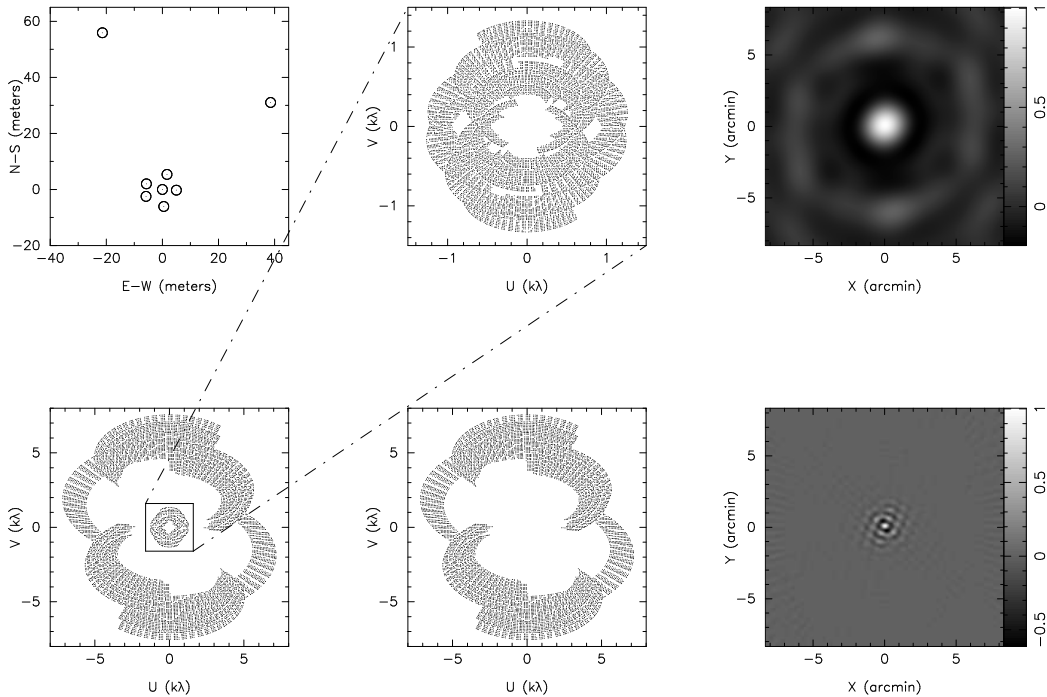}\\

\end{center}
\caption{{\bf Top left:} SZA telescope locations. {\bf Bottom left:} Resulting Fourier-space (\uv) coverage for a single track (all baselines). {\bf Top middle:} \uv\ coverage for the short-baselines only ($0-2~{\rm k\lambda}$). {\bf Bottom middle:} \uv\ coverage for the long baselines only ($2-8~{\rm k\lambda}$). 
{\bf Top right:} resulting point spread function (synthesized beam), short-baselines. {\bf Bottom right:} synthesized
beam, long baselines.}
\label{fig:fig1}
\end{figure}

Six of the SZA antennas are arranged in a close-packed configuration with
separations (baselines) ranging from $4.5-11.5~{\rm m}$ ($0.3-1.3~{\rm k\lambda}$ at
\ka), yielding high brightness sensitivity, and a resolution of $\sim 2
\arcmin$, for the detection
of the SZ effect in clusters
(see
Fig.~\ref{fig:fig1}). Two outer telescopes, one to the
north and one to the northeast of the central array, yield baselines
of up to $65~{\rm m}$ ($7.7~{\rm k\lambda}$ at \ka) to facilitate simultaneous
to optimize the surveying speed
detection of contaminating compact sources at a resolution of $\sim 0.3 \arcmin$.
 This hybrid configuration was chosen to optimize the surveying speed
of the SZA in an untargeted search.  Detailed imaging of the SZ effect can be achieved with alternate 
telescope configurations and with SZA observations at 90 GHz. 

The SZA centimeter-wave receivers use cryogenic $26-36~{\rm GHz}$ high electron mobility
transistor (HEMT) amplifiers \citep{pospieszalski1995}, with
characteristic receiver temperatures $T_{rx} \sim11-20~{\rm K}$.  Typical
system temperatures, including atmospheric contributions, range from
$\sim 30-60~{\rm K}$ in the 8-GHz band from $27$ to $35~{\rm GHz}$, used for these observations. Optical
fibers transport the signal (mixed down to intermediate frequencies of $1-9~{\rm GHz}$) to a
wideband hybrid correlator.  Sixteen 500 MHz-wide analog bands are
further divided into seventeen
$31.25~{\rm MHz}$ channels by 2-bit digital lag correlators, providing spectral discrimination as
well as high-sensitivity broadband measurements.
Additionally, the wide fractional bandwidth broadens the Fourier-space (\uv) coverage
appreciably, as is evident in Fig.~\ref{fig:fig1}.

\subsection{Observations}

Clusters were
typically observed for 10 hours about transit, which we 
refer to as a {\it track}.  Observations of the clusters
were interleaved with observations of a strong unresolved source every 15
minutes, to monitor variations in the instrumental gain.  \clA \ was
observed for a total of 15 tracks from August 23 to October 8, 2005, using
J1331+305 as its calibrator. \clB \ was observed for a total of 15
tracks between September 30 and October 28, 2005, also using J1331+305
as its calibrator.
\clC \ was observed for 3 tracks between
August 30 and September 4, 2005, using J1310+323 as its calibrator.

Data were calibrated and excised (flagged) in a data reduction package
specific to the SZA (see Section \ref{reductionAnalysis}).  Data in
this paper were taken during the commissioning period of the array,
during which $\sim 44\%$ of the data were flagged. Causes of flagging
were: an offline antenna ($17\%$); shadowing of one antenna by another
($15\%$); corruption of one 500-MHz band ($6\%$); lack of bracketing calibrator observations
($\sim5\%$), and rare, spurious correlations ($<1\%$). Subsequent
improvements to the instrument and observing strategy have
reduced flagged data to $< 5\%$ in good weather conditions.

In Table~\ref{obsTable}, we give the pointing center and cluster
redshift for each observation (taken from \citet{maughan2006}), along with details of the
observations, including the synthesized beam sizes (see \S\ref{sec:dataAnalysis})
for
both the short and long baselines.   We also present the achieved
\rms \ flux sensitivities for maps made with short and long-baseline data, and the corresponding brightness 
temperature sensitivities for the short-baseline maps.
The \rms \ in the short-baseline maps is expected to be marginally lower
than those of the long baselines (since there are 15 short baselines
and 13 long baselines).  For some of the observations of \clA\ and
\clB, however, one of the inner telescopes was offline, resulting in a larger
number of long baselines than of short baselines.

\begin{deluxetable}{lcrrccccccc}
\tabletypesize{\scriptsize}
\tablecolumns{8}
\setlength{\tabcolsep}{1mm}
\tablecaption{Cluster Observations}
\tablehead{
\colhead{Cluster Name}& \colhead{$z$\tablenotemark{a}}& \multicolumn{2}{c}
{\underline{Pointing Center (J2000)}}&\colhead{$\rm{t_{int}}$\tablenotemark{b}}& 
\colhead{$\rm{T_{sys}}$\tablenotemark{c}}& 
\multicolumn{3}{c}{\underline{Short Baselines (0-1.5k$\lambda$)} \hspace{2cm}}&
\multicolumn{2}{c}{\underline{Long Baselines (2-8k$\lambda$)}}\\

\colhead{} & \colhead{} & \colhead{$\alpha$}& \colhead{$\delta$} & \colhead{(hrs)} & 
\colhead{(K)} & \colhead{beam($\arcsec\times\arcsec \angle$)\tablenotemark{d}}& 
\colhead{$\sigma$(mJy)\tablenotemark{e}} & \colhead{B($\mu$K)\tablenotemark{f}} &
\colhead{beam($\arcsec\times\arcsec \angle$)\tablenotemark{d}}
& \colhead{$\sigma$(mJy)\tablenotemark{e}}}
\startdata
\clA  & 1.03& 14$^h$15$^m$11$^s$.2 &36$^{\circ}$12$^{\prime}$04$^{\prime\prime}$ & 34.1& 41.9 & 115.4$\times$131.2 -61.1 & 0.16 & 13.5 & 15.7$\times$21.4 87.2 & 0.16\\
\clB  & 0.92& 14$^h$29$^m$06$^s$.4 &42$^{\circ}$41$^{\prime}$10$^{\prime\prime}$ & 32.1& 41.7 & 109.9$\times$136.9 -60.9 & 0.17 & 13.6 & 15.6$\times$21.3 83.1 & 0.16\\
\clC  & 0.89& 12$^h$26$^m$58$^s$.0 &33$^{\circ}$32$^{\prime}$45$^{\prime\prime}$ & 7.6& 42.9  & 117.4$\times$125.4 -64.8 & 0.38 & 32.9 & 16.0$\times$21.2 79.2 & 0.37\\
\enddata

\label{obsTable}

\tablenotetext{a}{Redshifts from \cite{maughan2006}}
\tablenotetext{b}{On source integration time, unflagged data}
\tablenotetext{c}{Mean system temperature scaled to above the atmosphere}
\tablenotetext{d}{Synthesized beam FWHM and position angle measured from North through East}
\tablenotetext{e}{Achieved \rms \ noise in corresponding maps}
\tablenotetext{f}{Corresponding brightness sensitivity for the short baselines }
\end{deluxetable}

\subsection{Data Reduction and Calibration}
\label{reductionAnalysis}

The SZA data reduction 
package consists of a suite of MATLAB\footnote{The Mathworks, Version 7.0.4 (R14), \tt{http://www.mathworks.com/products/matlab}} routines, which
constitute a complete pipeline for flagging, calibrating, and
reducing visibility data (see \S\ref{sec:dataAnalysis}) before input to higher-level analysis
software. In the pipeline, data are converted to physical units and
corrected for instrumental phase and amplitude variations.
Calibration is performed using antenna-based gain solutions; data are flagged if
they do not meet designated criteria at each reduction step.  The pipeline outputs the calibrated, unflagged cluster visibility data.

The output of the correlator is a complex, dimensionless correlation
coefficient --- the ratio of correlated power to total power for each
baseline.  To convert these values to physical units (K), we first
calculate the system temperature of the antennas by comparison with
two loads of known temperature (a blackbody calibrator load on the
telescope, and the CMB).  System temperature measurements are made at the
beginning and end of every scan, where {\it scans} are defined as the
short (typically 5 to 15 minute) observations of either the cluster or
the calibrator within a track.  Data for a scan are written to disk in
consecutive $20$-second integrations, which is the native timescale on which calibration, flagging and Fourier-space calculations are performed.  Data with high system
temperatures (an indication of poor weather), or spurious calibration values (due
to readout error of the power sensors) are flagged at this
stage of the data reduction.

Absolute calibration is derived from observations of Mars, using fluxes predicted by the Rudy model \citep{rudy1987}.  Since Mars is
partially resolved on the longest baselines, a strong, unresolved source is
used to transfer the calibration from the short baselines.  This
absolute flux calibration is performed bimonthly.  Measured antenna
efficiencies (which include aperture efficiencies, atmospheric and system phase noise,
correlator efficiencies, and other factors) have fluctuated by
less than 5\% in the period for which the data in this paper were
taken. The absolute calibration was cross-checked by comparing 
SZA observations of Jupiter to those of WMAP and CBI \citep{wmapJupiter,cbiJupiter}.
Based on these measurements, we estimate the 
absolute flux calibration to be better than 10\% during these
observations.

Bandpasses are
measured at the beginning of every track using a flat-spectrum, unresolved radio
source with high signal-to-noise in each 31.25-MHz
channel. 
The first and last
31.25-MHz channel of each 500-MHz band are flagged, as they are
corrupted by aliasing in the lag correlator.
The bandpasses are corrected to remove differential gain (amplitude and phase)
across the channels within each 500-MHz band, and the absolute flux calibration on
Mars removes differential gain amplitude across the sixteen 500-MHz bands.  This
leads to a uniform gain calibration across the entire $8~{\rm GHz}$
bandwidth.

The phase from the changing geometric delay as a source is tracked
across the sky is removed in hardware by inducing a phase shift in the
local oscillator signal at each antenna, and digitally, in the lag
correlator, to account for the bandwidth. 
Accurate removal of the geometric
delay, which impacts the dynamic range of the instrument as well as the
accuracy to which source positions can be measured, requires precise
knowledge of each telescope location, as well as a measurement of the
displacement of the azimuth and elevation axes of the telescopes.  These locations have been determined to
$< 0.1~\rm{mm}$ by direct measurements and 
by observations of a set of strong radio sources across the
sky.

To remove residual instrumental phases, a strong unresolved source is observed every 15 minutes (over which
time the instrumental phase variations are typically much less than $35\degree$). The interpolated calibrator phase is removed on a
per-antenna basis; data are flagged if it is not possible to
interpolate the phase, either because a calibrator observation is bad
or because bracketing observations are missing. Data are also flagged if a
phase jump of more than $35\degree$ is observed between calibrator
pointings.  Observations of the phase calibrator are reused to perform
a relative amplitude calibration across a given track, to check for
any time-varying antenna gain. We have seen no indication that the
gain on an individual antenna changes significantly on timescales less
than 12 hours.  The gain amplitudes are scaled so that the
time-average of the gain over a track is the same for all antennas.

\section{Data Analysis}
\label{sec:dataAnalysis}

In the limit where sky curvature is negligible over the instrument's
field of view, the response of an interferometer on a single baseline,
known as a {\it visibility}, can be approximated by:
\begin{equation}
\label{vis2}
V(u,v)  = \int_{-\infty}^{+\infty}\int_{-\infty}^{+\infty}A_N(l,m)I(l,m) \times
\exp \{ -2\pi j [ul + vm]\}{dl\,dm}, 
\end{equation}
\normalsize where $u$ and $v$ are
the baseline lengths projected onto the sky, $l$ and $m$ are direction
cosines measured with respect to the $(u,v)$ axes,  $A_N(l,m)$ is the
normalized antenna beam pattern, and $I(l,m)$ is the sky intensity
distribution.  

\begin{figure}[Ht]

\centerline{
\includegraphics[scale=0.8]{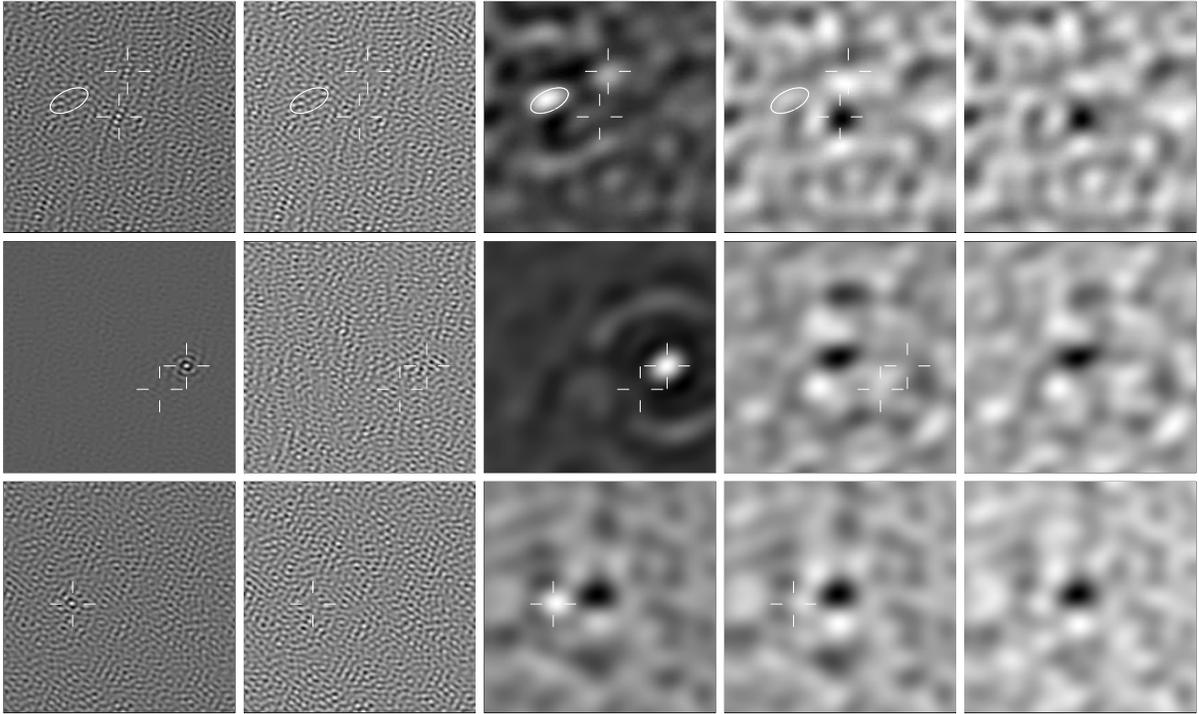}
}

\caption{Radio source subtraction in our three cluster
observations.  The rows, from top to bottom, depict observations of
\clA, \clB, and \clC.  The first column shows the dirty
map of the long-baseline data (\uv\ radii $>2~{\rm k\lambda}$).  The second
column shows the same maps, with fitted point sources
removed, as discussed in \S\ref{pt_src}, resulting in map residuals consistent with noise.  The locations of fitted point sources are indicated with cross-hairs. The third column shows the dirty map made from the short
baseline data (\uv\ radii $<2~{\rm k\lambda}$).  The fourth column shows the
same maps, following the removal of all sources
of radio emission.  In the top panel (\clA), we
have removed the extended emission from NGC 5529, indicated by an ellipse, resolved
out in the long-baseline map. 
The resulting cluster decrements reflect the shape of the synthesized beam
(see \S\ref{sec:dataAnalysis} and Fig.~\ref{fig:fig1}), convolved with the intrinsic cluster profile.   The final column shows the clean map
of the SZ decrement in  each cluster.  Images are $\approx21\arcmin$ on a side, nearly twice the FWHM of the primary beam.}
\label{fig:fig2}
\end{figure}

As implied by equation~(\ref{vis2}), an image of the source intensity
multiplied by the antenna beam pattern, also known as a {\it dirty
map}, can be recovered by Fourier transform of the visibility
data.  Note that in addition to modulation by the primary beam,
structure in the dirty map is convolved with a function that reflects
the incomplete Fourier-space sampling of a given observation.  This
filter function is the {\it synthesized beam} (see Figure
\ref{fig:fig1}), equivalent to the point-spread function for the
interferometer.  The unprocessed maps shown in the first four columns of Fig.~\ref{fig:fig2} are
examples of dirty maps.  
A $clean$ map, such as the ones in the last column of Fig.~\ref{fig:fig2}, is an image from which the full synthesized beam pattern has been
deconvolved, and the source model reconvolved with a Gaussian fit to
the central lobe of the synthesized beam.  Where a resolution
is quoted in this paper, it is this fitted Gaussian that is indicated.

All quantitative analyses described in this paper are the result of
simultaneously fitting models of the SZ cluster decrement and
sources of contaminating emission --- both point-like and extended ---
as detailed below. 
In all cases, the model is constructed in the sky-plane and
then multiplied by the primary beam across
the field of view.  
After performing a Fourier transform (as given by equation~(\ref{vis2})), the resulting model visibilities are
compared directly to the SZA data. In this way all fitting is done
in the Fourier-plane, where the noise characteristics
and the spatial filtering of the interferometer are well understood;
maps are used only for examination of the data and to identify cases where contaminating sources are
present.  The two distinct ranges in Fourier-space coverage
demonstrated in Figs.~\ref{fig:fig1} and \ref{fig:fig2} facilitate
removal of sources unresolved by the long-baseline synthesized beam
(point sources from the perspective of our instrument). Constraints on
the point-source flux come primarily from baselines including one or both of the outer telescopes, 
which probe small scales where the cluster signal is
exponentially damped, while the compact inner array provides
sensitivity to extended emission.

The frequency-dependent shape
of the primary beam used in the analysis is calculated
from the Fourier transform of the aperture illumination of each Cassegrain
telescope. The illumination is modeled
as a Gaussian taper, with a central
obscuration corresponding to the secondary mirror; the validity of this model has been confirmed by
holographic measurements of the primary mirrors. 

\begin{deluxetable}{lcrcrcccccl}
\tabletypesize{\scriptsize}
\tablecolumns{7}
\setlength{\tabcolsep}{0.9mm}
\tablecaption{Unresolved Radio Sources}
\tablehead{
\colhead{Cluster Field} &\colhead{$\#$} &\colhead{$RA$} & \colhead{$\sigma_{\rm RA}$} & \colhead{$DEC$}&\colhead{$\sigma_{\rm DEC}$} 
&\colhead{d\tablenotemark{a}}&\colhead{{\midka} Flux} & \colhead{$\alpha$\tablenotemark{b}}& 
\colhead{1.4 GHz flux\tablenotemark{c}} & \colhead{$\alpha$} \\
\colhead{} & \colhead{}& \colhead{(J2000)} &\colhead{(s)} 
 & \colhead{(J2000)}& \colhead{(\arcsec)} & \colhead{(\arcmin)}& \colhead{(mJy)}& \colhead{} & 
\colhead{(mJy)}& \colhead{(1.4/31 GHz)}}
\startdata
\clA  & 1 & $14^h15^m11^s.28$ &0.12& $+36^{\circ}12^{\prime}05^{\prime\prime}.1$ &1.2& 0.02 & $0.91\pm0.15$&& $3.15\pm 0.21$  & $0.40\pm 0.06$\\ 
      &2 & $14^h15^m07^s.43$ &0.13& $+36^{\circ}16^{\prime}17^{\prime\prime}.0$ &1.3& 4.28  & $1.01\pm0.17$&& $1.28\pm 0.16$  & $0.08\pm0.07$\\ 
\clB  & 1& $14^h28^m32^s.65$ &0.01& $+42^{\circ}40^{\prime}20^{\prime\prime}.8$ &0.1& 6.26  & $26.11\pm0.32$&$0.42\pm0.17$ & $43.6\pm 2.2$&$0.17\pm0.02$\\ 
      &2 & $14^h28^m46^s.14$ &0.15& $+42^{\circ}38^{\prime}11^{\prime\prime}.4$ &1.2& 4.77  & $1.38\pm0.21$&& $4.04\pm 0.24$  & $0.35\pm0.05$\\ 
\clC  & 1 & $12^h27^m18^s.63$ &0.08& $+33^{\circ}32^{\prime}06^{\prime\prime}.6$ &0.7& 4.35  & $4.56\pm0.42$& $1.1\pm1.4$& $23.2\pm1.2$& $0.53\pm0.03$\\ 
\enddata
\label{ptSrcTable}

\tablenotetext{a}{Distance from observation pointing center}
\tablenotetext{b}{Spectral index determined from SZA data alone (27-35 GHz)}
\tablenotetext{c}{Integrated FIRST flux at 1.4 GHz}

\end{deluxetable}

\subsection{Point Source Extraction}
\label{pt_src}

When fitting an unresolved radio source, hereafter referred to as a {\it point source}, we represent it 
by a delta function, parameterized by the intensity at 
the band center, $I_\midka$, and a spectral
index $\alpha$ over our sixteen $500~\rm{MHz}$-wide correlator
bands. The point source intensity at frequency $\nu$ is then:
\begin{equation}
\label{int_ps}
I_{ps}(l,m) = I_\midka \times \left(\frac{\nu}{\midka}\right)^{-\alpha} \delta(l -
l^{\prime})\,\delta(m - m^{\prime}),
\end{equation}
where $l^\prime$ and $m^\prime$ are the coordinates of the point source on
the sky. From equations~(\ref{vis2}) and (\ref{int_ps}), it can be seen
that the visibility amplitude due to a point source is simply its
intensity, weighted by the normalized primary beam response at the
source location.

The \clA \ and \clB \ cluster fields were each found to contain two point sources,
while only one point source was detected in the  \clC\ field
(see Table~\ref{ptSrcTable} and Fig.~\ref{fig:fig2}).  All detected
point sources have counterparts in the 1.4~GHz FIRST catalog
\citep{firstSurvey}. 
Spectral indices of sources detected with high signal-to-noise can be constrained using SZA data alone (see Table 2, where the limiting signal-to-noise is indicated by the poor constraint on the spectral index of the last source listed).
The spectral indices of weaker sources are not well constrained by
our data; for these sources the spectral index was fixed to the value
determined from the integrated flux at 1.4~GHz and the fitted total flux at \midka. 
The strong source in the \clB \ field provides a good test of our
ability to extract point sources. Following removal
of this source from fits to the long-baseline data alone, we obtain source
flux residuals of less than 3\% in both the long and short baseline maps.

\subsection{Extended Source Extraction}
\label{ext_src}

Where an extended source of emission is present in the field, its
frequency dependence is modeled as in equation~(\ref{int_ps}), with
spatial extent given by an elliptical Gaussian.  One such object, identified as the edge-on spiral galaxy NGC 5529,
was detected in the \clA \ field. The fitted source parameters are in
good agreement with the NVSS survey \citep{nvssSurvey}, which detects a source $\sim
2\arcmin$ in extent, at a position angle of $-64\degree$, with an
unresolved minor axis.  Comparison of the
integrated \midka \ flux for this object ($2.55\pm0.22$~mJy) with the
integrated flux from the NVSS catalog yields a spectral index
$\alpha_{\rm 1.4~GHz/\midka}=0.64\pm0.04$, consistent with 
synchrotron emission. Fig.~2 shows the effect of removing both 
point sources and this extended emission from the \clA \ cluster field.

\subsection{Cluster Parameter Estimation}
\label{markov}

\begin{figure}[t!]
\begin{center}
\includegraphics{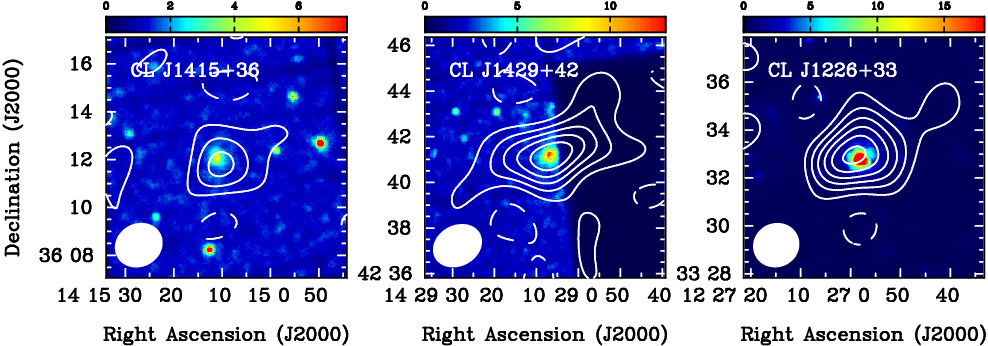}
\caption{Sunyaev-Zel'dovich Effect measurements (contours) overlaid 
on X-ray images (halftone) of \clA, \clB, and \clC. The contours are 
set at brightness levels corresponding to integer multiples of 
$\pm20~{\rm\mu K}$ for \clA, $\pm20~{\rm\mu K}$ for \clB \ and $\pm49~{\rm\mu K}$ for \clC, 
corresponding to $\pm 1.5$ times the map \rms, respectively (see Table~\ref{obsTable}); solid 
contours are negative and dashed contours are positive. In each panel,
the FWHM of the synthesized beam of the SZ observations is shown by 
the filled ellipse in the bottom left corner.
The X-ray images shown here are from the XMM EPIC/MOS instrument,
with an effective exposure time of
32~ks for \clA, 
66~ks for \clB, and 
45~ks for \clC.
The X-ray events are binned in $3\arcsec$ pixels,
and the images have been adaptively smoothed using a 
threshold of 70 differential
counts to determine the radii used in the smoothing kernel.}
\label{fig:fig3}
\end{center}
\end{figure}

In Fig.~\ref{fig:fig3}, we present clean maps of the 30 GHz emission of the three cluster fields
after the removal of the radio sources described above; the SZ decrement is
clearly detected for each cluster. Also shown in halftone is the 
corresponding X-ray emission for each field.
The radio
maps are for qualitative comparison only, illustrating the
confidence of the detections and the alignment of the SZ decrement
with the X-ray emission.  A quantitative analysis
of the SZ profile is rendered by fitting a Fourier-transformed model, described next,
to the visibility data.

We model the cluster gas density by
a spherical, isothermal \betaModel, described by
\begin{equation} 
n_e(r) = {n_{e}}_0 \left ( 1 + {r^2 \over r_c^2} \right)^{-3 \beta/2},
\label{elec_dens}
\end{equation} where the core radius $r_c$ and the power law index
$\beta$ are shape parameters, and ${n_{e}}_0$ is the central electron
number density.  
The model is a simple
parameterization of the gas density 
profile, traditionally used in fitting X-ray \citep[cf.][]{mohr1999a} and SZ data. Although more sophisticated parameterizations
have been shown to better reproduce details
of the density and temperature profiles, data taken with the SZA 
in its survey configuration lack the resolution
to constrain models with additional free parameters. 
Furthermore, this parameterization allows a direct 
comparison with the results of \citet{maughan2006}, who fit isothermal \betaModel s 
to the X-ray data for 
the three clusters
considered here.

The corresponding SZ temperature decrement is given by
\begin{equation} {\Delta T}\left ( \theta \right ) = \Delta T_0 \left( 1 +  {\theta^2 \over
\theta_c^2} \right)^{{1\over 2} -{3\beta \over 2}},\label{eq:DT/T}
\end{equation} 
where $\theta = r/D_A $, $\theta_c = r_c / D_A$, and $D_A$ is the angular
diameter distance.
The temperature decrement at zero projected radius, $\Delta T_0$, is related to
${n_e}_0$ by

\begin{equation}
\label{neo}
{n_e}_0 = \frac{\Delta T_0}{T_{\it CMB}} \frac{m_e c^2}{f(x)k_B\sigma_T}\frac{1}{T_e}\frac{1}{\sqrt{\pi} D_A \theta_c }\frac{\Gamma(\frac{3\beta}{2})}
{\Gamma(\frac{3\beta}{2}-\frac{1} {2})}.
\end{equation}

\begin{deluxetable}{lccc|cccc}
\tabletypesize{\footnotesize}
\tablecolumns{8}
\renewcommand{\arraystretch}{1.1}
\tablecaption{Resulting Fitted Model Parameters from SZA Observations}
\tablehead{
\colhead{Cluster Name} & \colhead{$\beta$} & \colhead{$\theta_c (\arcsec)$\tablenotemark{a}} & \colhead{scale (kpc/\arcsec)\tablenotemark{b}}
 & \colhead{${\Delta{\rm RA}}$ (\arcsec)} & \colhead{${\Delta{\rm DEC}}$ (\arcsec)} & 
\colhead{$\Delta T_0$ (mK)}& \colhead{$y_0$ ($10^{-4})$}}
\startdata
\clA    &  2/3 & 11.7   &  8.06 &$-26.8^{+14.8}_{-17.4}$ & $-10.8^{+14.1}_{-14.1}$& $-0.42\pm0.10$ & $0.80\pm 0.20$ \\
\clB    &  2/3 & 12.4   &  7.84 & $12.0^{+9.2}_{-10.5}$  & $-4.8^{+6.4}_{-6.7}$   & $-0.75\pm0.09$ & $1.44\pm 0.17$\\
\clC    &  2/3 & 14.6   &  7.77 &  $1.8^{+9.3}_{-8.7}$   & $16.8^{+7.1}_{-7.0}$   & $-1.68\pm0.17$ & $3.28\pm 0.33$\\
\enddata
\label{fitTable1}
\tablenotetext{a}{Adopted core radius from X-ray Measurements \citep{maughan2006}}
\tablenotetext{b}{Assuming $H_o = 70~{\rm km/s/Mpc}$, $\Omega_M = 0.3$, and $\Omega_\Lambda = 0.7$}
\end{deluxetable}

\begin{center}
\begin{deluxetable}{ccccc}
\tablewidth{4.5in}
\tabletypesize{\footnotesize}
\tablecolumns{5}
\renewcommand{\arraystretch}{1.1}
\tablecaption{Comparison of Fit Results for SZA and BIMA Observations of \clC}
\tablehead{
\colhead{Instrument} & \colhead{$\Delta{\rm RA}$ (\arcsec)} &
\colhead{$\Delta{\rm DEC}$ (\arcsec)} & 
\colhead{$\Delta T_0$ (mK)}& \colhead{$y_0$ ($10^{-4}$)}}
\startdata
SZA   &  $1.8^{+9.3}_{-8.7}$   & $16.8^{+7.1}_{-7.0}$ & $-1.68\pm0.17$ & $3.28\pm 0.33$\\
BIMA  &  $0.7^{+4.1}_{-3.9}$   & $12.8^{+4.3}_{-4.2}$ & $-1.75\pm0.16$ & $3.41\pm 0.31$\\
\enddata
\label{fitTable2}
\\Note: Fitted with fixed $\beta=2/3$ and $\theta_c=14\arcsec\!.6$
\end{deluxetable}
\end{center}

Best-fit values for the model parameters are determined using a Monte
Carlo Markov Chain analysis \citep[][and references therein]{bonamente2004,bonamente2006,laroque2006}.
The Markov chains are a sampling of the multi-dimensional likelihood for the
model parameters, given the SZ data; the histogram of values in the chain for each
parameter is thus an estimate of the probability distribution for that parameter, marginalized over the other model parameters.
In calculations where the fitted parameters ($\theta_c$, $\Delta T_0$)
are combined with other variables (such as $f_{gas}$ in the calculation of
electron temperatures described in \S~\ref{SZTemp}), each sample in the Markov
chain is paired with a random sample of those variables, to
obtain the probability distribution for the calculated quantity.
For any quantity determined from the Markov chains,
we quote the maximum-likelihood value, with an uncertainty obtained by integrating the
distribution for that quantity to a fixed probability density, until
68\% of the probability is enclosed.

In the analysis described in this and the following sections, two sets
of Markov chains were generated: one with a weak, uniform prior on
$\theta_c$ ($\theta_c \in [8\arcsec\!.5 - 30\arcsec]$), and one with a
strong prior ($\theta_c$ fixed at the value determined from X-ray
observations).  In both cases, the parameter $\beta$ was fixed to
 $2/3$, consistent with
X-ray and SZ observations of large cluster samples spanning a wide range in redshift
\citep[e.g.,][]{mohr1999a, laroque2006} as well as X-ray
observations of the clusters discussed here \citep{maughan2006}.

\label{sec:probability}

Table~\ref{fitTable1} presents cluster model parameters obtained
from the Markov chains.  The offset from
the pointing center of the best-fit cluster centroid is given in
columns 5 and 6.
For the isothermal \betaModel\ with
fixed $\beta$, although the total SZ fluxes
(and therefore cluster gas masses; see \S\ref{SZTemp})
are well constrained by
the SZA data alone, the data do not have sufficient resolution
to separately constrain $\Delta T_0$ and $\theta_c$.  
In Table~\ref{fitTable1} we therefore give constraints on the central decrement and central $y$
parameter for fixed $\beta = 2/3$, and 
for  $\theta_c$ fixed to the values from \citet{maughan2006}. For the masses presented in \S\ref{SZTemp}, however, we
present results derived from fits with only the weak prior on $\theta_c$.

In Table~\ref{fitTable2}, we compare model parameters from the $\sim
8$ hours of unflagged SZA observations with $\sim 42$ hours of BIMA
data \citep[see][]{joy2001} on \clC. Values for the cluster
locations, central decrements and $y$-parameters are
in excellent agreement. Note that the results agree
well even though the largest scale
probed by BIMA is roughly half that probed by the SZA.

\section{Results and Discussion}
\label{sec:results}

In this section, we use the Markov chains of model parameters
described in \S\ref{markov} to construct the probability distributions
of cluster properties, including the electron temperature, gas
mass and total mass.  We present values integrated to a
given overdensity radius, $R_{\Delta(z)}$, defined as the radius at
which the mean density of the cluster is related to the critical
density of the Universe by a fixed density contrast $\Delta(z)$.  
The density contrast $\Delta(z)$ is assumed to scale with redshift like 
the mean density of a virialized system, as determined from numerical simulations \citep{bryan1998}.

An estimate of the gas mass in the cluster can be obtained by
multiplying equation~(\ref{elec_dens}) by $\mu_e m_p$, the mass of the
proton weighted by the mean molecular weight of the electrons in the
gas, and integrating the result to the desired radius:
\begin{equation}
\label{eq:gasMassInt}
M_{gas}(R)  = {\mu_e m_p  {n_{e}}_0}\int_{0}^{R}\left ( 1 + {r^2 \over r_c^2} \right)^{-3 \beta/2}  4\pi r^2 dr.
\end{equation}
The central electron density ${n_e}_0$ is a function of the electron
temperature $T_e$ and the model parameters $\Delta T_0$, $\beta$ and
$\theta_{c}$, as given by equation~(\ref{neo}).

The total mass of the cluster can be estimated
by assuming hydrostatic equilibrium (hereafter HSE).  For the electron
distribution given by equation~(\ref{elec_dens}), this approximation
yields an analytic solution for the total cluster mass contained
within a radius $R$ of:
\begin{equation}\label{hseEquation}
\label{eq:hse}
M_{total}(R) = \frac{3k_BT_e\beta}{G\mu m_p} \frac{R^3}{r_c^2+R^2},
\end{equation}
where
$G$ is the gravitational constant, $\mu m_p$ is the mean
molecular mass of the gas, and $r_c$ is the core radius, related to
$\theta_c$ by the angular diameter distance.  We adopt a value of
0.3\,$Z_{\odot}$ for the cluster metallicity when calculating
both $\mu_e$ and $\mu$, consistent with X-ray observations of 
high-redshift clusters \citep{maughan2006}.

From equations (\ref{neo})--(\ref{eq:hse}), we see that
if we assume a value for the ratio of the gas mass to the total
cluster mass, hereafter referred to as the {\it gas-mass fraction},
$f_{gas}$, an estimate of electron temperature can be inferred, allowing the
masses to be determined without reference to an {\it a priori} value
for $T_e$. We employ this method below 
to obtain
cluster properties from the SZ data.  
For comparison, spectroscopically determined electron temperatures from X-ray measurements can be used to estimate the gas masses, total masses, and
$f_{gas}$ directly from the Markov chains.

\subsection{SZ Derived Cluster Properties}

\label{SZTemp}
Here we estimate cluster parameters by assuming a gas-mass fraction,
and using the SZ data to solve for $T_e$, following \citet{joy2001}
and \citet{laroque2003}. 
A previous study of a sample of 38 massive
clusters obtained a mean of $\overline{f}_{gas}=0.116\pm0.005$, from
masses evaluated within a radius of $R_{2500}$ (distinct from
$R_{2500(z)}$) \citep{laroque2006}.
In the calculation of the gas temperature for a single cluster, we therefore adopt a
Gaussian distribution of $f_{gas}$ with a mean of 0.116 and standard
deviation of 0.030, where we have scaled the reported error in the
mean by $\sqrt{37}$ to approximate the
measured distribution of gas-mass fractions.

We calculate the masses from the Markov chains obtained with a
uniform prior on the value of $\theta_c$
($\theta_c \in [8\arcsec\!.5 - 30\arcsec]$). For each entry in the chain, we sample the adopted $f_{gas}$ distribution and solve for  $T_e$
from equations~(\ref{neo})--(\ref{eq:hse}), evaluated at
$R_{2500}$ for consistency with \citet{laroque2006}.  
We use the resulting $T_e$ values to obtain estimates for $R_{2500(z)}$ and $R_{200(z)}$.  
The results, and the
gas and total masses calculated at the most likely value for the corresponding $R_\Delta(z)$ are presented in Table \ref{szRadiiTable}.  

For the ${\rm \Lambda CDM}$ cosmology used throughout this paper, the results
of numerical simulations suggest that the
density contrast of a cluster at the virial radius --- the boundary defined by the transition of the dynamical state
of the gas from infalling 
to hydrostatic equilibrium --- is approximately
${\rm \Delta\sim100}$ at $z=0$ \citep[][and references therein]{voigt2006}.
 For these clusters, total masses calculated at this density contrast (i.e., at $R_{100(z)}$) are
within 30\% of those calculated at $R_{200(z)}$.
The regions within the overdensity radii for which we quote results thus sample the cluster properties near
the cluster core ($R_{2500(z)}$) and near the virial radius
($R_{200(z)}$). For the three clusters presented in this paper, $R_{200(z)}$
corresponds to angular sizes on the order of 1.5\arcmin\ to 2.5\arcmin,
angular scales well sampled by the short-baseline data.

\begin{deluxetable}{ccc|cccc|cccc}
\tabletypesize{\scriptsize}
\tablecolumns{11}
\renewcommand{\arraystretch}{1.3}
\setlength{\tabcolsep}{0.75mm}
\tablecaption{Cluster Masses and ICM Properties Derived from SZ data}
\tablehead{
\colhead{} & \colhead{} &  \colhead{} & \multicolumn{4}{c}{\underline{Quantities within $R_{2500(z)}$ }} & \multicolumn{4}{c}{\underline{Quantities within $R_{200(z)}$}}\\
\colhead{Cluster Name}& \colhead{$\theta_c$} & 
\colhead{$T_e$} & \colhead{$R_{2500(z)}$} & \colhead{$M_{gas}$} & \colhead{$M_{total}$} & \colhead{$f_{gas}$} & \colhead{$R_{200(z)}$} & 
\colhead{$M_{gas}$} & \colhead{$M_{total}$} & \colhead{$f_{gas}$} \\

\colhead{}& \colhead{(\arcsec)} & \colhead{(keV)} & \colhead{(Mpc)} & \colhead{($10^{12}M_{\odot}$)} & \colhead{($10^{13}M_{\odot}$)} & \colhead{} & \colhead{(Mpc)} & \colhead{($10^{12}M_{\odot}$)} & \colhead{($10^{13}M_{\odot}$)} & } 
\startdata
\clA & [8.5, 30] & $3.7^{+0.8}_{-0.8}$  &  $0.17^{+0.03}_{-0.04}$ &  $3.4^{+0.7}_{-0.9}$  & $3.1^{+0.8}_{-0.8}$ & $0.114^{+0.024}_{-0.040}$ & $0.70^{+0.06}_{-0.09}$ &  $29.5^{+7.6}_{-4.6}$   & $18.6^{+4.5}_{-3.5}$ & $0.161^{+0.048}_{-0.051}$   \\
\clB & [8.5, 30] & $5.2^{+0.9}_{-0.7}$  &  $0.23^{+0.03}_{-0.04}$ &  $7.5^{+1.9}_{-1.4}$  & $7.3^{+1.6}_{-1.6}$ & $0.110^{+0.025}_{-0.036}$ & $0.88^{+0.09}_{-0.06}$ &  $55.9^{+6.8}_{-10.6}$  & $33.6^{+6.1}_{-4.7}$ & $0.142^{+0.052}_{-0.038}$ \\
\clC & [8.5, 30] & $8.3^{+1.8}_{-0.8}$  &  $0.31^{+0.03}_{-0.03}$ & $18.9^{+3.2}_{-3.4}$  &$16.8^{+3.7}_{-2.6}$ & $0.106^{+0.035}_{-0.027}$ & $1.14^{+0.10}_{-0.08}$ & $114.0^{+18.0}_{-18.0}$ & $71.9^{+13.3}_{-9.2}$ & $0.145^{+0.049}_{-0.040}$ \\  
\enddata
\label{szRadiiTable}
\end{deluxetable}

\subsection{Comparison to X-ray values}
\label{comp2x}

The three clusters discussed in this paper have been observed
with either the {\it Chandra} or {\it XMM-Newton} observatories,
allowing for a direct comparison of SZ derived properties with those
derived from X-ray data. 
In the previous section, we calculated masses to radii determined self-consistently from the SZ data; for a meaningful comparison with the X-ray results, however, these should be evaluated at the same physical radii.  In 
all comparisons of SZ and X-ray derived 
masses, we therefore calculate properties to the physical radii for which X-ray results 
have been reported, namely the $R^{X-ray}_{2500(z)}$ and $R^{X-ray}_{200(z)}$ values from
\citet{maughan2006}. 

In Table \ref{Tab:szTeMasses} we present the SZ-derived electron
temperature, gas masses, and total masses, with corresponding X-ray
determinations reproduced from \citet{maughan2006}.  We present values for two different priors on
$\theta_c$ (see \S\ref{markov}): the same weak prior used above, and a strong prior on $\theta_c$ to facilitate comparison with X-ray
derived properties.  Note that both priors give consistent results,
indicating that although SZA data alone cannot break the degeneracy of
$\theta_c$ with other model parameters ($\beta$ and $\Delta T_0$), it
can place strong constraints on cluster properties resulting from
combinations of these parameters.

\begin{deluxetable}{cccc|ccc|ccc}
\tabletypesize{\scriptsize}
\tablecolumns{10}
\renewcommand{\arraystretch}{1.3}
\setlength{\tabcolsep}{1.75mm}
\tablecaption{Comparison of SZ and X-ray Derived Temperature and Masses}
\tablehead{
\colhead{} & \colhead{} & \colhead{} & \colhead{} & \multicolumn{3}{c}{\underline{SZ Derived Quantities}} & 
\multicolumn{3}{c}{\underline{X-ray Masses}\tablenotemark{a}}\\
\colhead{Cluster Name} & \colhead{$R^{X-ray}_{\Delta(z)}$} & \colhead{$R$} & 
\colhead{$\theta_c$ \tablenotemark{b}}  & \colhead{$T_e$} & \colhead{$M_{gas}$} &  \colhead{$\it{M_{total}}$} & 
\colhead{$T_e$} & \colhead{$M_{gas}$}&\colhead{$\it{M_{total}}$} \\
\colhead{} & &\colhead{(Mpc)} &\colhead{($\arcsec$)} & \colhead{(keV)} & \colhead{($10^{12}M_{\odot}$)} & 
\colhead{($10^{13}M_{\odot}$)} & \colhead{(keV)} &\colhead{($10^{12}M_{\odot}$)} &\colhead{($10^{13}M_{\odot}$)}}
\startdata
\clA &  $\rm{R_{2500(z)}}$   & 0.23  & 11.7      & $3.7^{+0.8}_{-0.7}$ &   $6.2^{+1.5}_{-0.9}$  &  $5.6^{+1.3}_{-1.0}$  & $5.7^{+1.2}_{-0.7}$ & $6.7^{+1.2}_{-1.2}$    & $8.8^{+3.1}_{-2.5}$ \\
     &                       &       & [8.5, 30] & $3.7^{+0.8}_{-0.8}$ &   $5.6^{+1.8}_{-0.8}$  &  $5.1^{+1.0}_{-1.3}$  &                     &                        &                        \\
     &  $\rm{R_{200(z)}}$    & 0.88  & 11.7      & $3.7^{+0.8}_{-0.7}$ &  $38.2^{+9.2}_{-5.5}$  & $24.3^{+5.5}_{-4.3}$  & $5.7^{+1.2}_{-0.7}$ & $38.5^{+5.4}_{-4.3}$   & $38.3^{+12.0}_{-9.4}$ \\
     &                       &       & [8.5, 30] & $3.7^{+0.8}_{-0.8}$ &  $38.9^{+10.3}_{-6.3}$ & $24.0^{+5.1}_{-5.1}$  &                     &                        &                        \\
\clB &  $\rm{R_{2500(z)}}$   & 0.26  & 12.4      & $5.5^{+0.7}_{-1.0}$ &  $10.8^{+1.2}_{-2.0}$  &  $8.9^{+1.7}_{-1.1}$  & $6.2^{+1.5}_{-1.0}$ & $7.3^{+1.5}_{-1.6}$    & $10.5^{+4.8}_{-3.3}$ \\
     &                       &       & [8.5, 30] & $5.2^{+0.9}_{-0.7}$ &  $ 9.5^{+2.0}_{-1.7}$  &  $8.6^{+1.7}_{-1.8}$  &                     &                        &                        \\
     &  $\rm{R_{200(z)}}$    & 0.97  & 12.4      & $5.5^{+0.7}_{-1.0}$ &  $62.2^{+6.7}_{-11.2}$ & $39.7^{+4.9}_{-6.7}$  & $6.2^{+1.5}_{-1.0}$ & $42.9^{+7.6}_{-6.0}$   & $44.9^{+17.3}_{-1.29}$ \\
     &                       &       & [8.5, 30] & $5.2^{+0.9}_{-0.7}$ &  $62.7^{+8.7}_{-11.3}$ & $37.9^{+6.0}_{-6.0}$  &                     &                        &                        \\
\clC &  $\rm{R_{2500(z)}}$   & 0.35  & 14.6      & $8.9^{+1.2}_{-1.5}$ &  $23.7^{+3.6}_{-3.3}$  & $20.9^{+2.8}_{-3.2}$  & $10.6^{+1.1}_{-1.1}$& $21.5^{+1.9}_{-2.2}$   & $25.0^{+4.6}_{-4.3}$ \\
     &                       &       & [8.5, 30] & $8.3^{+1.8}_{-0.8}$ &  $23.7^{+3.0}_{-4.7}$  & $20.6^{+3.3}_{-3.7}$  &                     &                        &                        \\
     &  $\rm{R_{200(z)}}$    & 1.29  & 14.6      & $8.9^{+1.2}_{-1.5}$ & $130.2^{+19.6}_{-18.3}$& $84.7^{+11.5}_{-11.5}$& $10.6^{+1.1}_{-1.1}$& $119.0^{+8.9}_{-8.2}$  & $102.0^{+17.1}_{-16.8}$\\
     &                       &       & [8.5, 30] & $8.3^{+1.8}_{-0.8}$ & $130.4^{+21.1}_{-21.1}$& $80.3^{+14.6}_{-10.1}$&                     &                        &                        \\
\enddata
\label{Tab:szTeMasses}
\tablenotetext{a}{reproduced from \cite{maughan2006}}
\tablenotetext{b}{Prior on core radius, for SZ-derived quantities}
\end{deluxetable}

The SZ-derived electron temperatures --- and therefore total masses --- 
agree within the uncertainties with the X-ray values for the two highest-mass clusters.
The SZ and X-ray derived 
gas masses for all three
clusters are in good agreement. The SZ derived temperature for \clA, however, 
is marginally lower than the X-ray value. We do not believe the
discrepancy is significant, given the
confidence level for the detection of this cluster, 
uncertainties in the absolute calibration,
and possible systematic errors associated with the model assumptions.
We note, however, that the inconsistency would not be resolved by
assuming the SZ derived electron temperature is too low. If X-ray 
spectroscopic temperatures are adopted in place of the SZ derived temperatures, 
the resulting gas masses calculated from the SZ data would be marginally inconsistent with the
X-ray gas masses.

\section{Conclusion}
\label{sec:conclusion}

We present measurements and analyses of the Sunyaev-Zel'dovich (SZ) effect in three high-redshift
($0.89 \le z \le 1.03 $),
X-ray selected galaxy clusters. The observations were obtained
at 30 GHz during the commissioning period of the Sunyaev-Zel'dovich Array (SZA), an eight-element interferometer dedicated to SZ effect observations.  
The measurements are noteworthy in 
three respects: 1) they extend the redshift range of
reported SZ measurements to $z = 1.03$, 
2) they extend the low-mass range of reported SZ measurements 
down to $\sim 2 \times 10^{14}M_\odot$, and 3) with sensitivity to  
scales as large as $\sim 5'$, the SZA interferometric
observations provide sensitivity on angular scales
larger than the virial radii of the clusters.

Assuming isothermality and hydrostatic equilibrium for 
the intracluster medium, and
gas-mass fractions consistent with those derived from SZ effect 
and X-ray measurements at
moderate redshifts, we calculate the electron temperatures,
gas masses, and total masses of these clusters.
The SZ-derived total masses integrated to $R_{200(z)}$ are
$1.9 ^{+0.5}_{-0.4} \times 10^{14}M_\odot$
for \clA,
$3.4 ^{+0.6}_{-0.5} \times 10^{14}M_\odot$
for \clB, and 
$7.2 ^{+1.3}_{-0.9} \times 10^{14}M_\odot$
for ClJ1226.9\\+3332.
These values do not include
an overall calibration uncertainty ($<10$\%) and do not account for possible systematic uncertainties associated with the model assumptions.
A comparison of SZ derived properties to those derived using X-ray
data shows good agreement between the two methods.

The data presented here were taken in the SZA survey configuration.
Good agreement of
these measurements with prior X-ray results demonstrates the
capability of the SZA 
for probing high-redshift clusters, and bodes
well for upcoming SZ surveys.

\acknowledgments 

This paper is dedicated to the memory and spirit of James S.\ McDonnell.
We gratefully acknowledge and thank the James S.\ McDonnell Foundation for their
generous support of the SZA through a McDonnell Centennial Fellowship
to JEC. This support was critical to the development and construction
of the array. We also thank
the NSF Division of Astronomical Sciences for support through
grants AST-0096913 and AST-0604982 to the construction and continued operation
of the SZA, and we thank the University of Chicago for its generous
matching funds to the initial NSF-AST grant. We thank 
the Kavli Institute of Cosmological Physics (KICP) for support
provided by its NSF Physics Frontier Center grant PHY-0114422 to expand the
SZA from six to eight telescopes and to outfit the array with 90 GHz
receivers. We are also grateful for the intellectual support of the
KICP's members and
visitors. We acknowledge the David and Lucile Packard Foundation 
for support
of the precursor SZ research using the OVRO and BIMA arrays, as well as
partial support for the SZA construction. 
This work is also supported in part by NSF AST-0507545 and NSF AST-05-07161
awarded to Columbia.  CG, SM,
and MS acknowledge support from NSF Graduate Research Fellowships. MR acknowledges support from a Fermi Fellowship from
the Enrico Fermi Institute at the University of Chicago. AM
gratefully acknowledges the generous support of the Alfred P.\ Sloan
Foundation. SM thanks R. Rodriguez, M. Teyssier and J. Donovan for insightful discussions.
We thank 
the staff of the Owens Valley Radio
Observatory for their outstanding support, and the Radio Astronomy
Laboratories at U.C.\ Berkeley and UMASS for providing critical components. 
Lastly, we thank M.~Bonamente, B.~Maughan, R.~Plambeck, and A.~Stanford for 
their helpful discussions and suggestions on this work.

\bibliographystyle{apj}

\bibliography{ms}
\end{document}